# Symmetry breaking driven by petahertz electron acceleration in a centrosymmetric organic superconductor

Y. Kawakami[1], T. Amano[1], H. Ohashi[1], H. Itoh[1], Y. Nakamura[2], H. Kishida[2], T. Sasaki[3], G. Kawaguchi[4], H. M. Yamamoto[4], K. Yamamoto[5], S. Ishihara[1], K. Yonemitsu[6], and S. Iwai[1]*

[1]Department of Physics, Tohoku University, Sendai 980-8578, Japan

[2]Department of Applied Physics, Nagoya University, Nagoya 464-8603 Japan

[3]Institute for Materials Research, Tohoku University, Sendai 980-8577, Japan

[4]Institute for Molecular Science, Okazaki 444-8585, Japan

[5]Department of Applied Physics, Okayama Science University, Okayama, 700-0005 Japan

[6]Department of Physics, Chuo University, Tokyo 112-8551, Japan

71.27.+a, 74.25.Gz, 78.47.J-

*s-iwai@tohoku.ac.jp

Abstract

Charge acceleration during an intense light field application to solids attracts much attention as elementary processes in high-harmonic generation and photoelectron emission [1-7]. For manipulating such attosecond dynamics of charge, carrier-envelope-phase (CEP:relative phase between carrier oscillation of light field and its envelope function) control has been employed in insulators, nanometal and graphene [8-10]. In superconducting materials, collective control of charge motion is expected because of its strongly coherent nature of quasi-particles.

Here, in a layered organic superconductor, second harmonic generation (SHG) is observed by using a single-cycle 6 femtosecond near infrared pulse, which is in contrast to the common belief that even harmonics are forbidden in the centrosymmetric system. The SHG shows a CEP sensitive nature and an enhancement near the superconducting temperature. The result and its quantum many-body analysis indicate that a polarized current is induced by non-dissipative acceleration of charge, which is amplified by superconducting fluctuations. This will lead to petahertz functions of superconductors and of strongly correlated systems.

An electromagnetic oscillation of light cannot directly access a spatial symmetry breaking because of its symmetric nature on the time axis (i. e., the time average of the oscillation is zero). However, recent developments of ultrashort laser technologies enable us to control the direction of charge motion by carrier-envelope phase (CEP) control of a strong light field [8-10]. Considering non-perturbative and non-dissipative light-matter interactions during an ultrashort pulse, we can expect petahertz control of the spatial inversion symmetry in solids.

In the non-dissipative regime (before electron scatterings occur substantially), a current $j(t)$ driven by a light field $E(t) = E_0(t)\sin(\omega t - \varphi_{\text{CEP}})$ [$E_0$(t): envelope of a single-cycle pulse, $\varphi_{\text{CEP}}$: CEP] [Fig. 1(a)] is given by $j(t) \propto \int_0^t E(t)dt$ [Fig. 1(b)], indicating that an induced current can be modulated by the CEP. This is in contrast to the fact that a net current driven by a light-field is zero in the dissipative regime $j(t) = \sigma E(t)$, where $\sigma$ is conductivity. Such a non-dissipative current can break the symmetry of a charge density in the momentum space [or equivalently a breaking of the spatial inversion symmetry in the sense that the induced current is no longer described as an odd function of $E(t)$ of the moment], resulting in current induced SHG [11-13, for conventional (dissipative-)current-induced SHG]. Thus, the SHG (and the spatial inversion symmetry breaking) is induced even in the centrosymmetric system, if the non-dissipative current is driven by light.

Our target material is a layered organic superconductor κ-(BEDT-TTF)$_2$Cu[N(CN)$_2$]Br [14-18] with a transition temperature $T_{\text{SC}}$=11.6

K (Fig. 1(c)). Superconducting fluctuations above $T_{SC}$ ($T_{SC} < T \lesssim 2 T_{SC}$) have been discussed in analogy with a pseudogap in high-$T_{SC}$ superconducting cuprates in the temperature-$t/U_{dimer}$ phase diagram [17, 18], where $t/U_{dimer}$ is the ratio of an inter-dimer transfer integral $t$ to the effective on-site Coulomb energy for a dimer $U_{dimer}$, as shown in Fig. 1(c). In superconductors, optical responses have been discussed in terms of non-equilibrium quasi-particle dynamics on the time scale of picosecond [19-22], coherent excitation of the Higgs mode [23] and light-induced superconductivity [24, 25]. On the other hand, petahertz light functions driven by charge dynamics during a light pulse is also expected, reflecting the large energy scale of electron correlation (ca. 1 eV) [26].

In this letter, we report SHG for a strong light field (10 MV/cm=1 V/nm) in a centrosymmetric organic superconductor κ-(BEDT-TTF)$_2$Cu[N(CN)$_2$]Br (single crystal). A CEP dependence of the SHG shows that it is induced by the non-dissipative current. Superconducting fluctuations above $T_{SC}$ amplify the SHG.

Figure 2(a) shows spectra (measured at 6 K) of SHG and third harmonic generation (THG) with peak energies of 1.5 eV [SHG, red ($E_{Fund}||c$, $E_{SH}||c$) and green ($E_{Fund}||a$, $E_{SH}||c$, x 0.73) curves] and 2.2 eV [THG, blue curve ($E_{Fund}||c$, $E_{SH}||c$, x 0.024] for the fundamental photon energy of 0.75 eV. Here, $E_{Fund}$, $E_{SH}$, and $E_{TH}$ indicate the electric fields of the fundamental light, SHG and THG, respectively. Note that the SHG is originally forbidden in this centrosymmetric system (orthorhombic with $P$nma symmetry [27, 28]) in a conventional perturbation theory.

As shown in the upper panel of Fig. 2(b), the SHG is polarized parallel to the c-axis ($E_{SH}||c$) for both excitation polarizations [$E_{Fund}||c$ (red line), $E_{Fund}||a$ (green line)], although the THG shows the usual polarization which is the same as that of the fundamental pulse (lower panel). Such unusual polarization dependence of the SHG is discussed below. That is not attributed to a surface SHG [29], because the result above ($E_{Fund}||a$, $E_{SH}||c$) is confirmed by rotating the sample not to depend on the s- or p-polarized configuration.

Figure 2(c) shows the temperature dependence of the peak intensities for SHG [$I_{SH}$: closed red circles ($E_{fund}||a$, $E_{SH}||c$), open red circles ($E_{fund}||c$, $E_{SH}||c$)] and THG [$I_{TH}$: blue] (normalized by the intensities at 60 K), indicating that the SHG grows up toward $T_{SC}$, although $I_{TH}$ does not depend on the temperature. The fluctuation of $I_{SH}$ for ($E_{fund}||c$, $E_{SH}||c$) is larger than that for ($E_{fund}||a$, $E_{SH}||c$) because of stray lights from the intense/broad THG. The spectral shape is almost independent of the temperature, showing that the net intensity of the SHG is increased near $T_{SC}$.

Considering that the pulse width of 6 fs is shorter than the scattering time [ca. 40 fs = h/(0.1 eV)] in organic conductors, a possible mechanism for the unconventional SHG is the non-dissipative current discussed above. Furthermore, the temperature dependence [Fig. 2(c)] shows that the non-dissipative current is enhanced by the superconducting fluctuations.

We can demonstrate relevance of this scenario by measuring the CEP dependence of the SHG, because the non-dissipative current is sensitive to the CEP as mentioned above (Fig. 1). Figure 3(d) shows the intensity of the SHG as a function of the relative CEP ($\Delta\varphi_{CEP}$). During a period of the CEP, the SHG shows two maxima at around $\Delta\varphi_{CEP}=1/2\pi$ and $3/2\pi$. Considering that the directions of the light-induced current cannot be distinguished by the SHG measurement, i.e., that currents with a same amplitude and opposite directions give same SHG intensities, the above $\Delta\varphi_{CEP}$ dependence is quite reasonable. This CEP dependence of the SHG is completely consistent with the fact that the unconventional SHG is attributed to the non-dissipative light-induced current.

To clarify the origin of the SHG more in detail, we theoretically calculate the current density $j$ in a two-dimensional three-quarter-filled Hubbard model for a 98x98-site system in the framework of the time-dependent Hartree-Fock approximation. The details are described in supplementary 1, where the emergence of SHG is also checked by the exact diagonalization method for a 16-site system. The calculated SHG and THG spectra ($\omega J$, $J$ denotes the absolute value of the Fourier transform of $j$ [6, 7]) for $E_{SH}||c$, $E_{TH}||c$ with electric field amplitudes ($F$ [V/angstrom]) of 0.16, 0.06, and 0.006 ($\hbar\omega=0.7$ eV) are respectively shown in Figs. 4(a)-4(c). $I_{SH}$ is sensitive to the CEP [Fig.4(d)], which is consistent with the result shown in Fig. 3(d). On the other hand, the observed anisotropy of the SHG [Fig. 2(b)] cannot be reproduced by the theory, i. e, $I_{SH}$ (theory) shows the polarization that is parallel to the fundamental polarization for both $E_{fund}||c$ and $E_{fund}||a$. For reproducing the observed polarization dependence, $j||c$ should be driven for both $E_{fund}||c$ and $E_{fund}||a$, which does not contradict a point group analysis of $\chi^{(2)}$ tensor for the orthorhombic structure [class mm2($C_{2v}$) (after due consideration of the symmetry breaking uniaxially induced by the current $j$)] [supplementary 2]. In fact, the results of transient reflectivity measurements are consistent with the above results of the SHG, i.e., the responses of $E_{pr}||\mathbf{c}$ are larger than $E_{pr}||\mathbf{a}$ for both $E_{pu}||\mathbf{a}$ and $E_{pu}||\mathbf{c}$ ($E_{pu}$ and $E_{pr}$ are the electric fields of pump- and probe- lights, respectively) [supplementary 3]. This result shows that $E_{fund}||\mathbf{a}$ induces charge motion along the **c**-axis [supplementary 2].

As to the calculated excitation intensity ($I$, defined as $F^2$) dependences [Figs. 4(e) (SHG) and 4(f) (THG) for electrons in the HOMOs], we notice that $I_{SH} \propto I^{2.9}$ for $I > 10^{-4}$ and $I_{TH} \propto I^{3.1}$ for $I > 10^{-6}$ as shown by the green dashed lines. Above the "threshold" in the theory [supplementary 4], $I_{SH} \propto I^{2.9}$ does not agree with the experimental result $I_{SH} \propto I^{2.1}$ [inset of Fig. 2(a)]. It would be caused by the fact that the intra-molecular charge motion is not taken into account, though the intra-dimer charge motion is fully taken into account. In fact, the Fourier intensity of the time profile of the charge density in the HOMO of a molecule (to which the intra-molecular optical transition is sensitive) shows a square dependence, which is consistent with our experimental result. [supplementary 5].

Another important issue is that the spectral bandwidth of the SHG (130 meV) in Fig. 2(a) is narrower than that of the 6-fs excitation pulse [almost equal to the Fourier limit of the pulse (500 meV)]. The 0.6-2 eV is known as a "spectral window" in the organic superconductors, i.e., we have no clear reflectivity and absorption bands for ||**c** polarization. Therefore, a spectral deformation of the SHG ($E_{SH}||\mathbf{c}$) owing to an absorption loss is not the

reason of the narrow bandwidth [supplementary 6]. Considering that the band width of the THG (430 meV) is close to that of the 6-fs excitation pulse, the coherence of the SHG survives ca. 30 fs after the light-field application. Note that the wavelength dispersion of the phase matching condition of SHG for the reflection configuration is as small as that of THG. The narrow bandwidth of the SHG is not attributed to that.

The coherence time of the SHG (30 fs) is comparable to the electronic scattering time of ~40 fs (=h/(0.1 eV)) in organic conductors, indicating that the coherence time is governed by the electronic scattering [supplementary7]. This is consistent with the fact that the bandwidth of the SHG is independent of temperature. On the other hand, the intensity is sensitive to superconducting fluctuations, indicating that such non-dissipative charge motion is enhanced by the superconducting fluctuations. The non-dissipative charge motion on the petahertz time scale indicates that the microscopic mechanism of the superconductivity in this system is related to the Coulomb interaction of ca. 1 eV. This is a very important perspective for petahertz functions and attosecond science of superconductors and of strongly correlated systems in addition to already realized terahertz functions of them [23-24].

In summary, this letter demonstrates that SHG is induced by the non-dissipative current in the centrosymmetric organic superconductor κ-(BEDT-TTF)$_2$Cu[N(CN)$_2$]Br. The SHG is amplified by superconducting fluctuations near $T_{SC}$. A narrow bandwidth of 130 meV shows that the coherence of the SHG survives for 30 fs after the 6 fs light-field application.

Methods

**Sample preparation.** A single crystal and a thin film of κ-(BEDT-TTF)$_2$Cu[N(CN)$_2$]Br (single crystal 0.7 × 0.5 × 0.8 mm for axes *a*, *b*, *c* was prepared using the methods described in previous studies [26].

**6 fs infrared pulse generation.** The 6 fs pulse is generated by the method described in Refs. 26, 30, i. e., a broadband infrared spectrum covering 1.2–2.3 μm is obtained by focusing a carrier-envelope phase (CEP) stabilized idler pulse (1.7 μm) from an optical parametric amplifier (Quantronix HE-TOPAS pumped by Spectra-Physics Spitfire-Ace) onto a hollow fibre set within a Kr-filled chamber (Femtolasers). Pulse compression is performed using both active mirror (OKO, 19-ch linear MMDM) and chirped mirror (Femtolasers and Sigma-Koki) techniques.

**Measurement and CEP control of SHG.**

 We performed SHG and THG measurements for the single crystal using a 6 fs pulse with a reflection geometry (incident angle is smaller than 3-degree). The intensity and polarization of the fundamental (excitation range :0.01 to 2 mJ/cm$^2$) pulse are controlled by a pair of wire-grid CaF$_2$ polarizers[30]. The SHG and the THG are detected by an photomulipler tube (Hamamatsu R13456) after passing through a spectrometer (JASCO, M10). The CEP of the fundamental pulse is controlled by a pair of glass plates with the incidence angle of θ (Fig. 3(a)) and detected by the 2f-3f interferometer (2f and 3f are generated using β-BBO ) [Figs. 3(b) and 3(c)].

Data Availability

The data that support the plots within this paper and other findings of this study are available from the corresponding author upon reasonable request.

Figure Legends

Fig. 1 **CEP control of non-dissipative current and phase diagram of κ- type BEDT-TTF salts**

**a** Light-field $E(t) = E_0(t)\sin(\omega t - \varphi_{\text{CEP}})$, where $\varphi_{\text{CEP}}$ =0 (red), $1/4\pi$ (orange), $1/2\pi$ (green), $\pi$ (blue) [$E_0(t)$: envelope of a single-cycle pulse]. **b** Non-dissipative light induced current $j(t) \propto \int_0^t E(t)dt$. **c** Temperature-$t/U_{\text{dimer}}$(band width) phase diagram of κ-(BEDT-TTF)$_2$X, which is extracted based on controlling the chemical pressure (=$t/U_{\text{dimer}}$).

Fig. 2 **SHG and THG spectra of κ-(BEDT-TTF)$_2$Cu[N(CN)$_2$]Br, and their polarization and temperature dependences**

**a** The red and green lines indicate the SHG spectra of single crystalline κ-(BEDT-TTF)$_2$Cu[N(CN)$_2$]Br (6 K) for the polarizations of ($E_{\text{fund}}$||**c**, $E_{\text{SH}}$||**c**) (red line) and ($E_{\text{fund}}$||**a**, $E_{\text{SH}}$||**c**) (green line, x 0.73). The blue line shows the THG spectrum (x 0.024) for ($E_{\text{fund}}$||**c**, $E_{\text{TH}}$||**c**). The intensity of the fundamental light ($I_{\text{fund}}= E_{\text{fund}}^2$) is 2 mJ/cm$^2$. Inset shows $I_{\text{fund}}$ dependences of $I_{\text{SH}}$ (red line for $E_{\text{fund}}$||**c**, $E_{\text{SH}}$||**c**) and $I_{\text{TH}}$ (blue line for $E_{\text{fund}}$||**c**, $E_{\text{TH}}$||**c**). **b** Polarization dependences of $I_{\text{SH}}$ (upper panel) and $I_{\text{TH}}$ (lower panel) for $E_{\text{fund}}$ ||**c** (red line) and $E_{\text{fund}}$ ||**a** (green line), respectively. **c** Temperature dependences of $I_{\text{SH}}$ (closed red circles: $E_{\text{fund}}$||**a**, $E_{\text{SH}}$||**c**, open red circles: $E_{\text{fund}}$||**c**, $E_{\text{SH}}$||**c**) and $I_{\text{TH}}$ (closed blue circles, $E_{\text{fund}}$||**c**, $E_{\text{TH}}$||**c**). Both are normalized by the respective intensities at 60 K.

Fig. 3 **CEP control of SHG**

**a** Schematic illustrations of the CEP ($\varphi_{CEP}$=0, $\pi$) and the pair of BK-7 plates (thickness 1 mm) for changing the relative CEP $\Delta\varphi_{CEP}$ ($\theta$: incident angle of light). **b** Interference spectrum between 2f and 3f of the fundamental light. The 2f and 3f are generated using $\beta$-BBO. **c** $\Delta\varphi$ as a function of $\theta$ (which is obtained by the 2f-3f interferometer). **d** Intensity change of the SHG ($\Delta I_{SH}/I_{SH}$) as a function of $\Delta\varphi$. The red curve is obtained by averaging the data (guide to the eye).

Fig. 4 **Calculated spectra of SHG and THG**

**a**, **b**, **c** Calculated spectra of $\omega J$ showing SHG and THG for light-field (which is polarized parallel to the **c**-axis) $F$= 0.16(**a**), 0.06 (**b**), and 0.006 (**c**), respectively. **d** CEP dependence of the calculated SHG intensity [peak intensities of $\omega J$ at 1.4 eV (SHG)] for $F$=0.1. **e**, **f** Calculated intensities of SHG (**e**) and THG (**f**) as a function of $I=F^2$. The green dashed lines in e and f indicate $I^{2.9}$(**e**) and $I^{3.1}$(**f**)

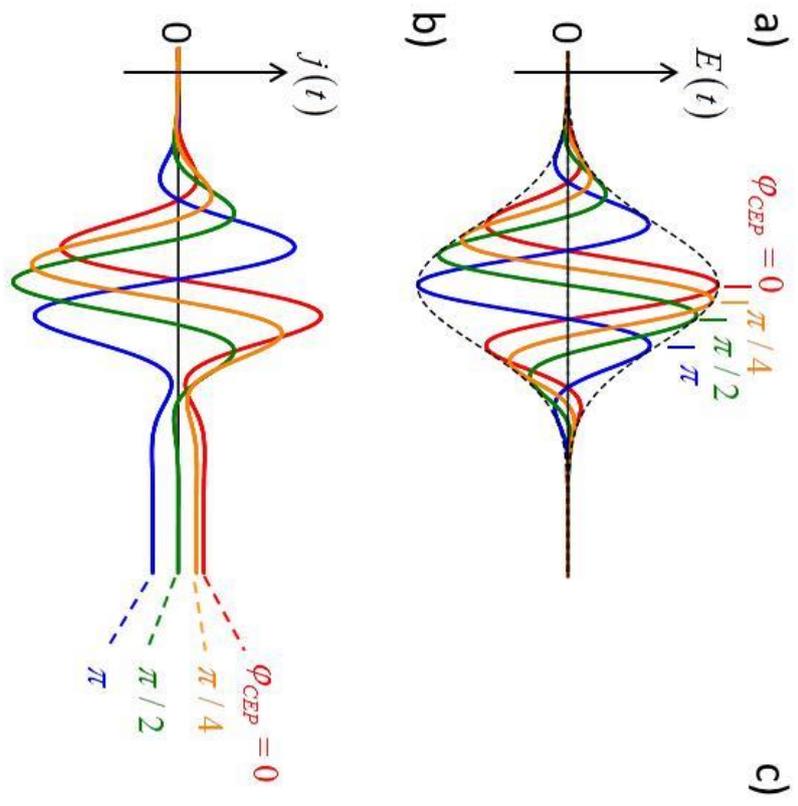
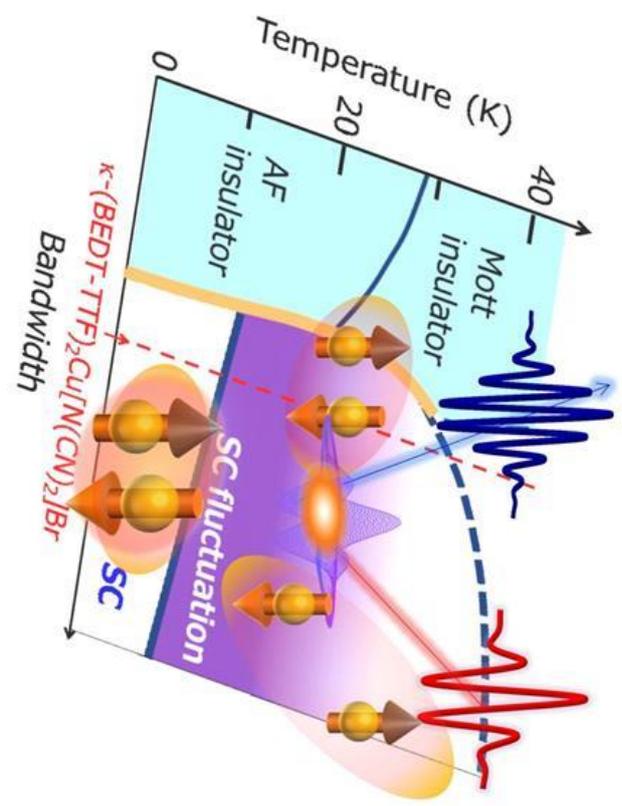

Fig. 1

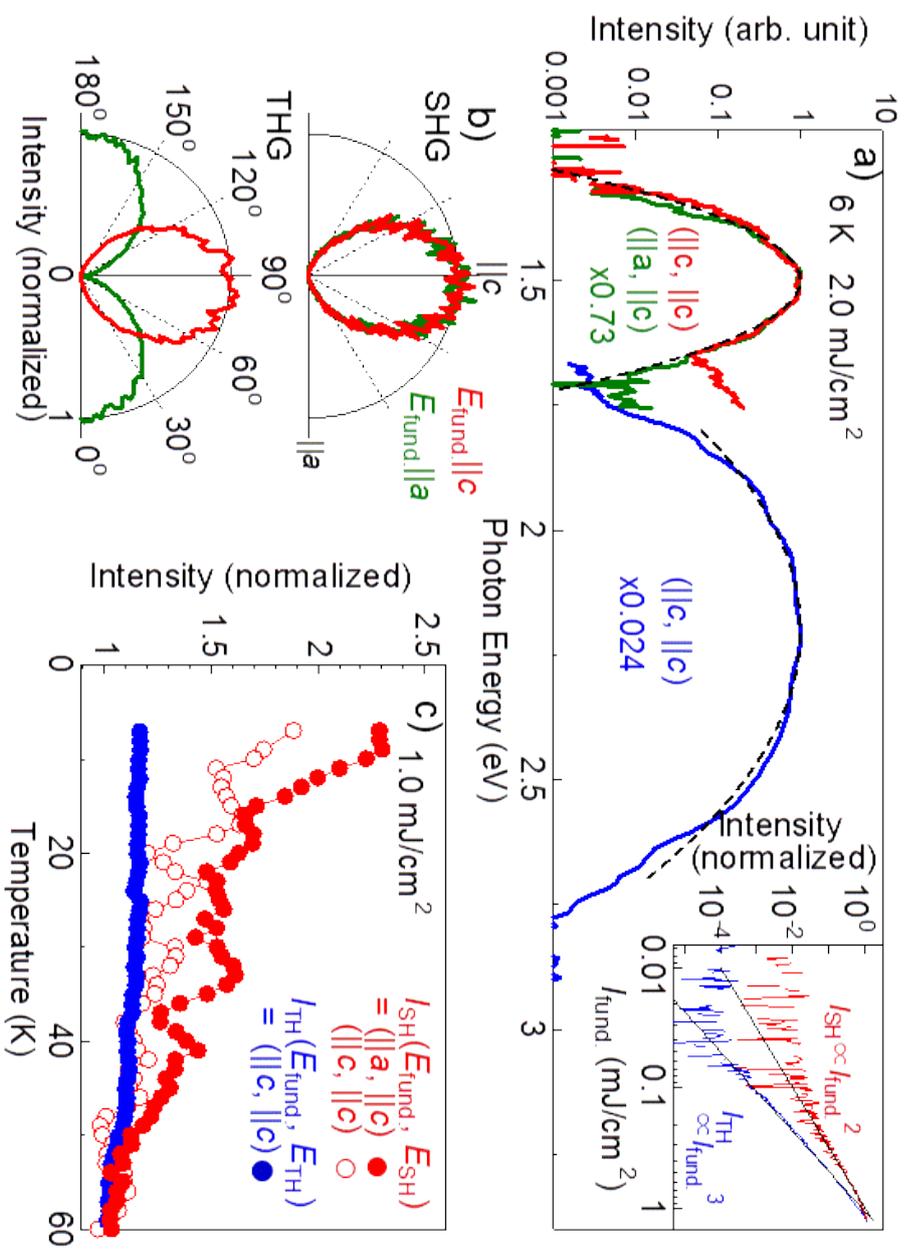

Fig. 2

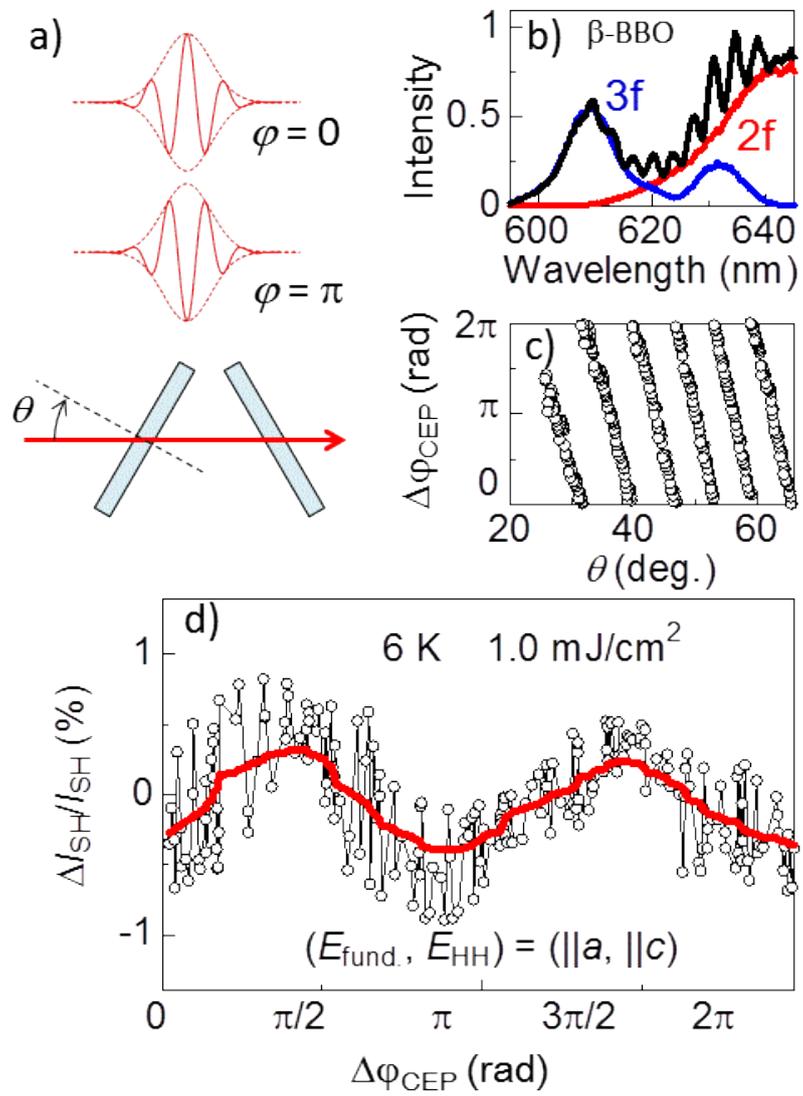

Fig. 3

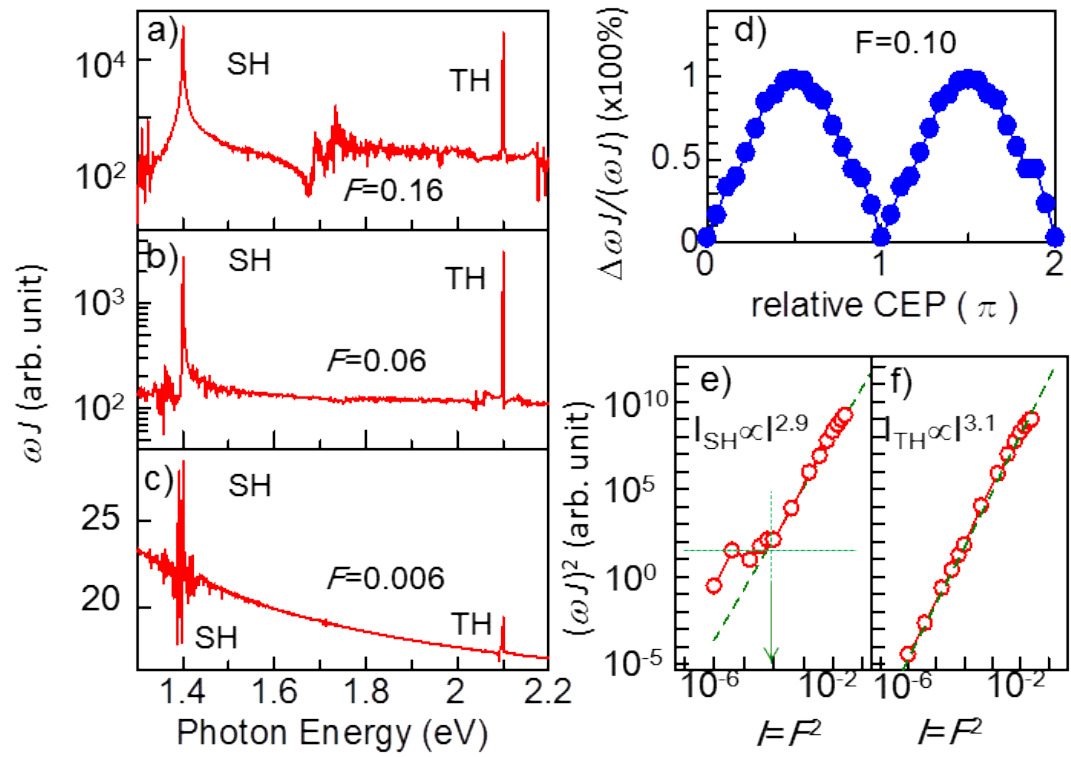

Fig. 4

Supplementary 1

Technical details of theoretical consideration (2D Hubbard model)

We use the Hubbard model at three-quarter filling, $H = \sum_{\langle i,j \rangle \sigma} t_{ij}(c_{i\sigma}^\dagger c_{j\sigma} + c_{j\sigma}^\dagger c_{i\sigma}) + U \sum_i n_{i\uparrow} n_{i\downarrow}$, where $c_{i\sigma}^\dagger$ creates an electron in the highest occupied molecular orbital (HOMO) with spin $\sigma$ at site $i$, and $n_{i\sigma} = c_{i\sigma}^\dagger c_{i\sigma}$. The parameter $U$ represents the on-site Coulomb repulsion and $U$=0.8 eV is used. The transfer integral $t_{ij}$ depends on the bond $ij$. The molecular arrangement is taken from the structural data, from which $t_{ij}$ are estimated with the extended Hückel method [27, 28]. The initial state is the Hartree-Fock ground state. Photoexcitation is introduced through the substitution $c_{i\sigma}^\dagger c_{j\sigma} \rightarrow \exp\left[\frac{ie}{\hbar c} \mathbf{r}_{ij} \cdot \mathbf{A}(t)\right] c_{i\sigma}^\dagger c_{j\sigma}$ with relative intermolecular position $\mathbf{r}_{ij} = \mathbf{r}_j - \mathbf{r}_i$. We use the vector potential $\mathbf{A}(t) = \theta(t)\frac{F}{\omega_{\text{fund}}}[\cos(\omega_{\text{fund}} t - \phi) - \cos\phi]$, which corresponds to $\mathbf{E}(t) = \theta(t) F \sin(\omega_{\text{fund}} t - \phi)$, with fundamental photon energy $\hbar\omega_{\text{fund}}$=0.7 eV. The time-dependent Schrödinger equation is numerically solved. We calculate the Fourier transform of the current density $\mathbf{j}(t) = -\langle \frac{\partial H}{N \partial \mathbf{A}} \rangle$ with $N$ being the number of unit cells for 500 cycles. The absolute value of its Fourier transform is denoted by $J$. The SHG and THG are evaluated as $\omega J$ (the absolute value of the Fourier transform of $d\mathbf{j}/dt$) at $\omega$=2$\omega_{\text{fund}}$ and $\omega$=3$\omega_{\text{fund}}$ [6, 7], respectively. To check the emergence of SHG, we also calculate $J$ by using the exact diagonalization method for a 16-site system during 10-cycle irradiation [26, S1]. Fourier spectra of $\omega J$ are shown by the closed blue circles in Fig. s1 [(a) $F$=0.16, (b) $F$=0.06, (c) $F$=0.006]. As shown in Figs. s1(a) and s1(b), the SHG peak is confirmed.

In Figs. 4(a)-4(c), the SHG is overestimated, because of the following reason. When we fully take electron correlations into account by employing the exact diagonalization method, we find that inter-site repulsive interactions substantially reduce the coherence of the charge oscillations that are responsible for the SHG, just as they reduce the coherence of the nonlinear charge oscillations that are responsible for the stimulated emission [S1]. On the other hand, we also find that they do not affect the THG so much. In the main text, we employ the time-dependent Hartree-Fock approximation, which cannot treat such decoherence effects, so that the SHG is overestimated and thus comparable with the THG.

Supplementary 2

Anisotropy of SHG

The SHG is polarized to the **c**-axis ($E_{SH}||c$) for both fundamental polarizations ($E_{Fund}||c$, $E_{Fund}||a$) as shown in Fig. 2(b), although the THG shows the usual polarization ($E_{TH}||c$ for, $E_{fund}||c$ and $E_{TH}||a$ for $E_{fund}||a$). As described in the main text, the observed anisotropy of the SHG ($E_{SH}||c$ for $E_{Fund}||a$) cannot be reproduced by the theory which takes only the HOMO for each BEDT-TTF molecule into account, i. e., ($E_{fund}||c$, $E_{SH}||c$) and ($E_{fund}||a$, $E_{SH}||a$) in the calculation.

The polarization dependence of the SHG in κ-(BEDT-TTF)$_2$Cu[N(CN)$_2$]Br should be described by the $\chi^{(2)}$ tensor of the orthorhombic structure [class mm2 (C$_{2v}$) after due consideration of the symmetry breaking uniaxially induced by the current $j$],

$$\chi^{(2)} = \begin{pmatrix} 0 & 0 & 0 & 0 & \chi^{(2)}_{xzx} & \chi^{(2)}_{xxy} \\ 0 & 0 & 0 & \chi^{(2)}_{yyz} & 0 & 0 \\ \chi^{(2)}_{zxx} & \chi^{(2)}_{zyy} & \chi^{(2)}_{zzz} & 0 & 0 & 0 \end{pmatrix}$$

, where $\chi^{(2)}_{zxx} = \chi^{(2)}_{xzx}$, $\chi^{(2)}_{yyz} = \chi^{(2)}_{zyy}$

The relation ($E_{\text{fund}}||c$, $E_{\text{SH}}||c$) is easy to understand because $\chi^{(2)}_{zzz} = \chi^{(2)}_{ccc}$ is nonzero for $j||c$ (Uniaxial symmetry breaking is assumed in the direction of the c-axis). On the other hand, because the above $\chi^{(2)}$ tensor of the orthorhombic structure [class mm2 ($C_{2v}$)] is based on the uniaxial symmetry breaking in the direction of the c-axis, the observed anisotropy of the SHG ($E_{\text{fund}}||a$, $E_{\text{SH}}||c$) cannot be understood from $\chi^{(2)}_{zxx} = \chi^{(2)}_{caa}$ above.

In contrast to the assumption, the light-field is actually applied along the a-axis ($E_{\text{fund}}||a$). In fact, the observed anisotropy ($E_{\text{fund}}||a$, $E_{\text{SH}}||c$) indicates that $j||c$ should be induced by $E_{\text{fund}}||a$. Meanwhile, the results of transient reflectivity measurements are consistent with the above results, i.e., the optical responses measured by a transient reflectivity (pump-probe) are larger for $E_{\text{pr}}||c$ than for $E_{\text{pr}}||a$ for both $E_{\text{pu}}||c$ and $E_{\text{pu}}||a$ ($E_{\text{pu}}$ and $E_{\text{pr}}$ are the electric fields of the pump- and probe- lights) [supplementary 3]. Thus, charge motion along the c-axis is actually driven by $E_{\text{pu}}||a$ as well.

The microscopic reason for the induction of $j||c$ by $E_{\text{fund}}||a$ remains unclear. However, it is reasonable to consider that the induction of $j||c$ by $E_{\text{fund}}||a$ is caused by the intra-molecular charge transfers, which are beyond the scope of the theory, because the frequency of the SHG ($E_{\text{SH}}||c$) is resonant to the weak intra-molecular transition ($E||c$) of BEDT-TTF molecules as shown in the inset of Fig. s5(b) [supplementary 6]. As shown in supplementary 6], the intra-molecular transition is almost polarized to the **a**-axis because of the tilting of BEDT-TTF molecules toward the **a**-axis. The induction of $j$||**c** by $E_{\text{fund}}$||**a** is related to the intra- molecular charge motion in addition to the inter-molecular charge motion.

The intra-molecular transitions cause an effect similar to hole doping on the HOMO band. The pseudogap behavior is observed only under hole doping and is caused by the van Hove singularity in the density of states [S2]. Under photoirradiation, the situation is different from the hole doping in equilibrium. However, the HOMO band is transiently less than three-quarter filling, and the one-electron states near the van Hove singularity are responsible for the transient charge-disproportionate state. Thus, the transient state is considered to be largely affected by the van Hove singularity, which is located on the Z point [S2] corresponding to the momentum along the c-axis. Because the carriers in the HOMO band would possess momenta mainly along the c-axis, it is natural for charge to oscillate along the c-axis.

## Supplementary 3

### Polarization analysis of transient reflectivity (pump-probe) measurement

To demonstrate a response of $j$||**c** ($E_{\text{pr}}$||**c**) under the excitation polarization of $E_{\text{pu}}$||**a**, we investigate the polarization dependence of a transient reflectivity (pump-probe) measurement. Figure s2 shows time evolutions of reflectivity changes ($\Delta R/R$) at 0.62 eV (excitation is made by 6 fs, 1 mJ/cm$^2$ ,

0.6-0.9 eV). The $\Delta R/R$ for $E_{pr}||c$ [red ($E_{pu}||c$) and green ($E_{pu}||a$) curves] are much larger than those for $E_{pr}||a$ [blue ($E_{pu}||c$) and magenta ($E_{pu}||a$) curves)]. (In reference 26, only the results for $E_{pu}||c$, $E_{pr}||c$ have been discussed). Therefore, it is clear that light-induced charge motion occurs easily along the **c**-axis for both $E_{pu}||a$ and $E_{pu}||c$. The similar polarization dependence has also been observed in κ-(BEDT-TTF)$_2$Cu[N(CN)$_2$]Cl (insulating phase) even under weaker excitation condition of 3.1 eV (probe energy=1.55 eV )[S3].

Supplementary-4

Calculated SHG spectra below and above the threshold

We have clear difference in calculated spectra of the SHG below and above the threshold, i. e., a sharp peak at 1.4 eV [=0.7 eV (fundamental photon energy in the theory) x2] is seen above the threshold [Fig. s3(a)], whereas we notice an oscillating structure on the background of the fundamental component below the threshold [resulting in the fluctuation in Fig. 4(e)] [Fig. s3(b)]. However, the experimental threshold of the SHG is not clear because of detection limits. [inset of Fig. 2(a)].

Supplementary-5

Discrepancy between the experimental and the theory on excitation intensity dependence

$I_{SH} \propto I^{2.9}$ [Fig. 4(e)] does not agree with the experimental result $I_{SH} \propto I^{2.1}$ [inset of Fig. 2(a)]. Although the reason of this discrepancy remains unclear, it can be related to the fact that the intra-molecular charge motion is not taken into account, as mentioned in the main text. In fact, the Fourier intensity (the square of the absolute value of the Fourier transform) of the time profile of the charge density in the HOMO of a molecule to which the intra-molecular optical transition is sensitive shows square dependence above the threshold as follows. The closed blue circles in Fig. s4 show the Fourier intensities $\rho^2$ for the charge density during 500-cycle irradiation as for $(\omega J)^2$, and the black circles show those during 50-cycle irradiation. The red circles show the calculated SHG intensities $(\omega J)^2$ [red circles in Fig. 4(e)]. The blue, black and red dashed lines indicate $I^{2.0}$ (blue and black) and $I^{2.9}$ (red). Thus, the discrepancy between the observation ($I_{SH} \propto I^{2.1}$) and the theory ($I_{SH} \propto I^{2.9}$) would be due to the intra-molecular charge redistribution.

**Supplementary-6**

**Transmittance spectrum of κ-(BEDT-TTF)$_2$Cu[N(CN)$_2$]Br thin film in near infrared - visible region.**

The energy range (0.6-2 eV) is known as a "spectral window" between inter-molecular (< 0.5 eV) and intra-molecular (> 2.3 eV) charge transfer bands in the organic superconductors [S4-S6], i.e., we have no large reflectivity or absorption bands except for a peak at 1.5 eV as shown in references [S7, S8]. We measure a transmittance spectrum of a thin film [S9] to investigate the influence of the absorption to the SHG, because it is difficult to estimate an absorption coefficient by Kramers-Kronig analysis of a reflectivity in such a "transparent" spectral region.

Figures s5(b)(c) show absorption spectra of a κ-(BEDT-TTF)$_2$Cu[N(CN)$_2$]Br thin film with a thickness of 150 nm[S9] for *E*||**c** (b) and *E*||**a** (c) (*E* is the electric field of light) [Fig. s5(a) shows the SHG and THG (linear scale)]. The peak structure at 1.37 eV for *E*||**a** is attributed to the intra-molecular transition of a BEDT-TTF molecule. The anisotropy of this intra-molecular transition, i. e. the peak for *E*||**a** is much larger than that for *E*||**c**, is associated with the crystal structure of κ-(BEDT-TTF)$_2$Cu[N(CN)$_2$]Br, where the long axis of BEDT-TTF molecules is tilted towards the **a**-axis [27, 28, S10]. The small and broad spectral feature around 1.5 eV for *E*||**c** [inset of Fig. s5(b)] does not affect the spectral shapes of the SHG for both (*E*$_{fund}$||**c**, *E*$_{SH}$||**c**) and (*E*$_{fund}$||a, *E*$_{SH}$||**c**) [Fig. 2(a)]. Thus, a spectral deformation of the SHG (*E*$_{SH}$||**c**) owing to an absorption loss is not the reason of the narrow bandwidth.

Supplementary 7

Coherence time of non-linear charge motion as studied by transient reflectivity measurement with double pump pulses

The electronic coherence after the field application is also seen in another strong light field effect in this compound [26]. We perform a transient reflectivity (ΔR/R) measurement with a double-pump pulse to make an insight of coherence induced by a strong-light field. This method has been employed for elucidating electronic and vibrational coherence [S11-S13]. Here, the coherence of the non-linear charge motion, which has been assigned as the origin of stimulated emission[26], is discussed.

Figures s6(a)(b) show an interferogram pattern which is detected by ΔR/R measurement with double pump pulses [as a function of Δt [=time difference between two pump pluses; -20-110 fs (a), -2-17 fs (b)]. It is probed at $t_d$ [=time delay between the 1st pump and the probe (0.62 eV= peak of stimulated emission) pulses]= 50 fs. An interferogram reflecting the coherence of the non-linear charge motion [26] [red curves in Figs. s6 (a) and s6(b)] indicates that electronic coherence survives ca. 70 fs after the pulse [A black curve in (b) shows the autocorrelation of the pulse]. The Fourier spectrum of the interferogram [red curve in Fig. s6(c)] is analogous to the spectrum of the stimulated emission reflecting the non-linear charge motion observed in the single-pumped measurement [26] [Fig. s6(d)].

This result confirms that the spectral band width of ΔR/R in the single-pumped measurement is determined by the coherence time of the non-linear charge motion. Thus, the electronic coherence of the non-linear charge motion can survive after the pulse. The coherence time of the stimulated emission (ca. 70 fs) and the SHG (30 fs) are comparable to the electronic scattering time of ~40 fs (=h/(0.1 eV)) in organic conductors.

**Figure legend of supplementary**

**Fig. s1 Calculated spectra of SHG (exact diagonalization)**

**a, b, c** Calculated spectra of $\omega J$ showing SHG and THG by the closed blue circles ($E_{\text{fund}}||\mathbf{c}, j||\mathbf{c}$) for $F$= 0.16(**a**), 0.06 (**b**), and 0.006 (**c**), respectively. The spectra of $\omega J$ calculated by the time-dependent Hartree Fock approximation [Figs. 4(a), 4(b) and 4(c)] are shown by the red curves after normalization. (supplementary 1)

**Fig. s2 Polarization analysis of transient reflectivity (pump-probe) measurement**

Time evolutions of transient reflectivity change [$\Delta R/R$) at 0.62 eV (excitation by 6 fs, 1 mJ/cm$^2$, 0.6-0.9 eV pulse, 6 K]. The time profiles of $\Delta R/R$ for $E_{\text{pr}}||\mathbf{c}$ are shown by the red ($E_{\text{pu}}||\mathbf{c}$) and green ($E_{\text{pu}}||\mathbf{a}$) curves. Those for $E_{\text{pr}}||\mathbf{a}$ are shown by the blue ($E_{\text{pu}}||\mathbf{c}$) and magenta ($E_{\text{pu}}||\mathbf{a}$) curves. (supplementary 2, 3)

**Fig. s3 Calculated SHG spectra below and above the threshold**

Calculated spectra of the SHG above the threshold (**a** $F$=0.06) and below the threshold (**b** $F$=0.006). (supplementary 4)

**Fig. s4 Charge density in the HOMO**

Fourier intensities of the time profile of the charge density in the HOMO of a molecule $\rho^2$ as a function of $I=F^2$ (closed blue circles for those during 500-cycle irradiation and black circles for those during 50-cycle irradiation).

The blue, black and red dashed lines indicate $I^{2.0}$ (blue and black) and $I^{2.9}$ (red). The red circles show the calculated SHG intensities [$(\omega J)^2$] [red circles, Fig. 4(e)] . (supplementary 5)

Fig. s5 **Transmittance spectrum of κ-(BEDT-TTF)$_2$Cu[N(CN)$_2$]Br thin film**
**(a)** SHG (red and green curves) and THG (blue curve) spectra (same as Fig. 2(a), but shown by a linear scale). **(b)(c)** absorption spectra of a κ-(BEDT-TTF)$_2$Cu[N(CN)$_2$]Br thin film with a thickness of 150 nm [S9] for $E||c$ (b) and $E||a$(c) ($E$ is the electric field of light). Inset in (b) shows an enlarged view of the intramolecular transition ($E||c$) and the SHG. (supplementary 6)

Fig. s6 **Coherence time of the the nonlinear charge oscillation**
**a, b** Interferogram pattern which is detected by a transient reflectivity (ΔR/R) measurement with double pump pulses [as a function of Δt (=time difference between two pump pluses) covering the range of -20-110 fs (**a**), -2-17 fs (**b**)]. It is probed at $t_d$ [=time delay between the 1st pump and the probe (0.62 eV= peak of stimulated emission[26]) pulses]= 50 fs. The black line in **b** shows autocorrelation of the 6 fs pump pulse. **c** Fourier spectra of the interferogram (red curve) and the autocorrelation of the pulse (black curve). **d** Spectra of stimulated emission which are measured by the single-pump transient reflectivity measurement ($t_d$=10 fs(red curve), 200 fs(blue curve))[26]. (supplementary 7)

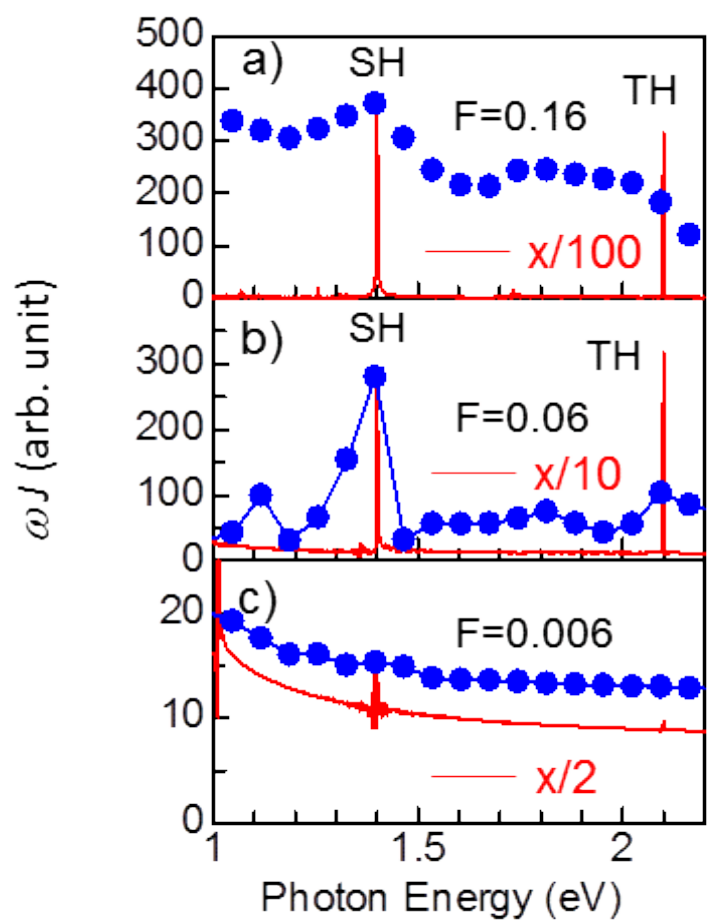

Fig. s1

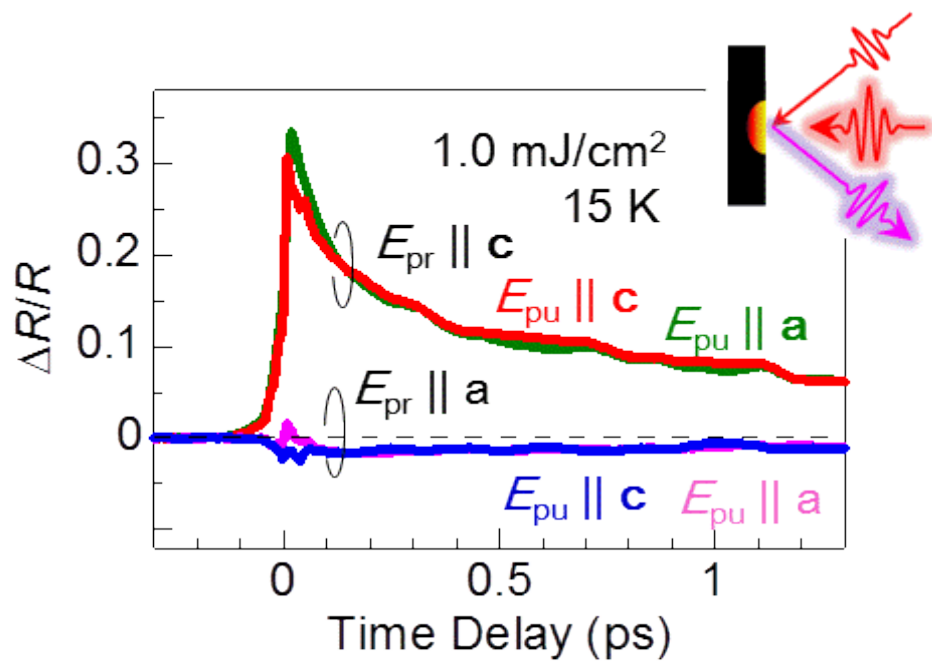

Fig. s2

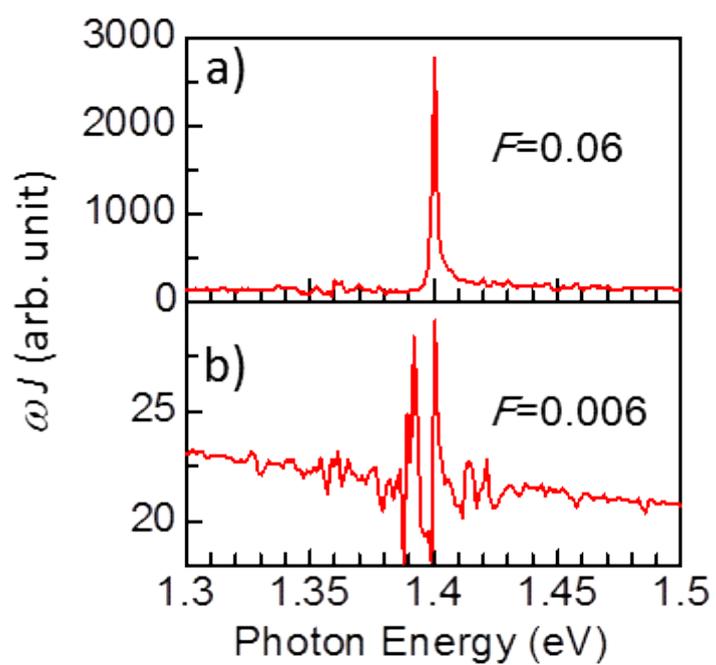

Fig. s3

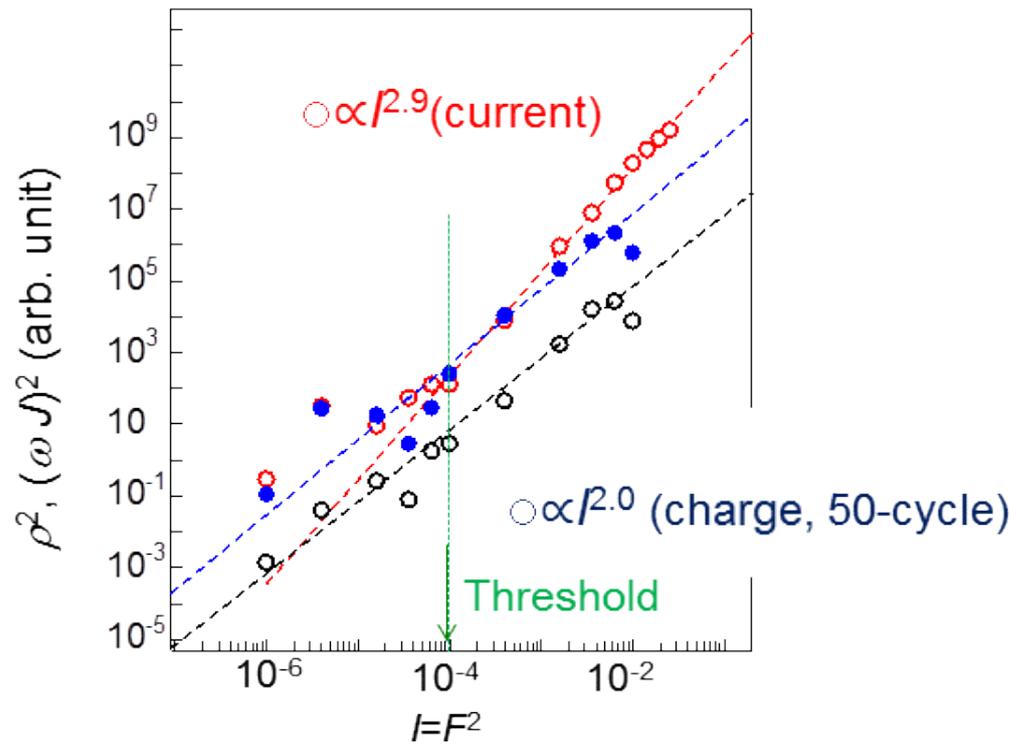

Fig. s4

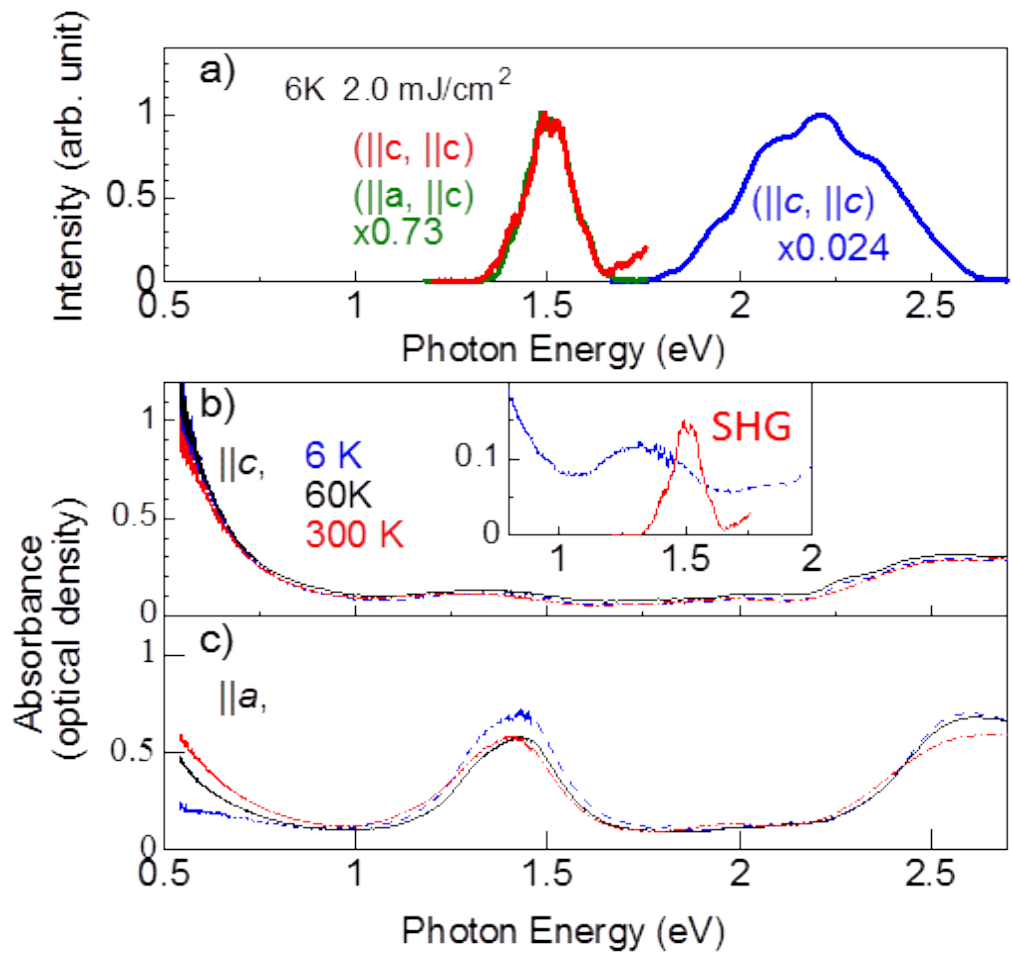

Fig. s5

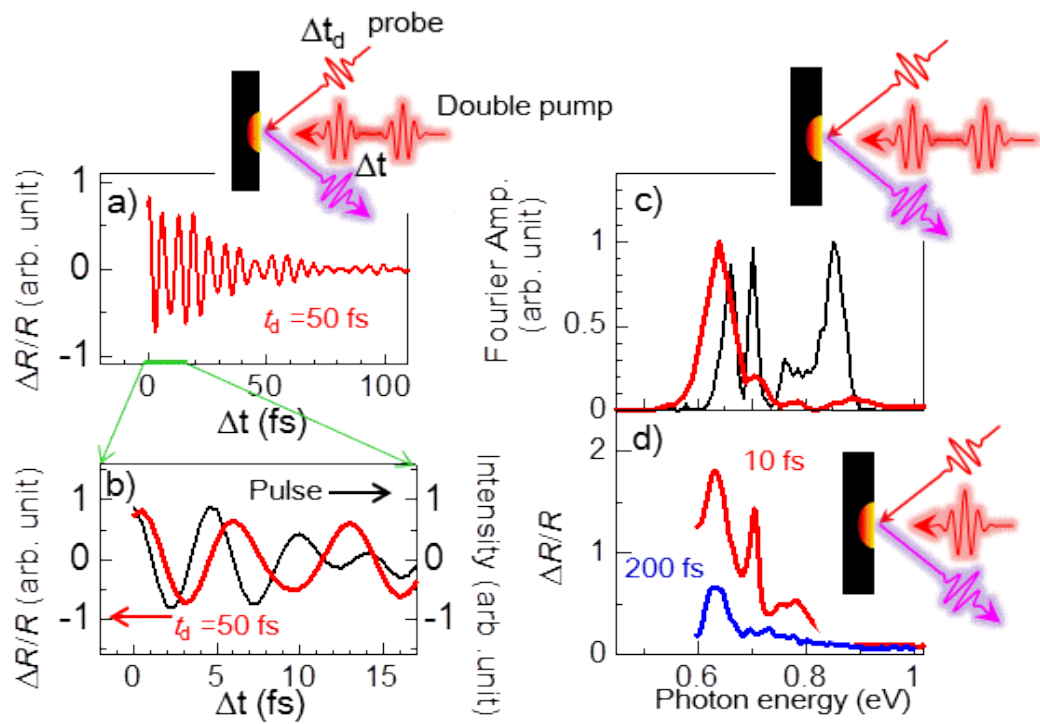

Fig. s6